\documentclass[letterpaper]{article} 
\usepackage{aaai25}  
\usepackage{times}  
\usepackage{helvet}  
\usepackage{courier}  
\usepackage[hyphens]{url}  
\usepackage{graphicx} 
\usepackage{natbib}  
\usepackage{caption} 
\usepackage{algorithm}
\usepackage{algorithmic}

\usepackage{newfloat}
\usepackage{listings}
\DeclareCaptionStyle{ruled}{labelfont=normalfont,labelsep=colon,strut=off} 
\floatstyle{ruled}
\newfloat{listing}{tb}{lst}{}
\floatname{listing}{Listing}
\usepackage{amsmath}
\usepackage{amssymb}
\usepackage{caption}
\usepackage{amsthm}
\usepackage{enumitem}
\usepackage{subcaption}
\usepackage{tikz}
\usepackage{pgfplots} 
\pgfplotsset{compat=1.16} 
\usepackage{bigstrut,array,multirow,tabularx}
\usepackage{caption}
\usepackage{pifont}
\usepackage{dsfont}
\usepackage{mathtools}

\usepackage{algorithm}
\usepackage{algorithmic}
\usepackage{makecell} 
\newtheorem{theorem}{Theorem}

\usepackage{booktabs}

\usepackage{pgfplotstable}
\usepackage{multicol}
\nocopyright 

\title{CF-KAN: Kolmogorov-Arnold Network-based Collaborative Filtering \\ to Mitigate Catastrophic Forgetting in Recommender Systems}
\author{
    $\text{Jin-Duk Park}^{\rm 1}, \text{Kyung-Min Kim}^{\rm 2}, \text{Won-Yong Shin}^{\rm 1}$
}
\affiliations{
    \textsuperscript{\rm 1} Yonsei University\\
    \textsuperscript{\rm 2} NAVER\\

}

\usepackage{bibentry}

\begin{document}

\maketitle

\begin{abstract}
Collaborative filtering (CF) remains essential in recommender systems, leveraging user--item interactions to provide personalized recommendations. Meanwhile, a number of CF techniques have evolved into sophisticated model architectures based on multi-layer perceptrons (MLPs). However, MLPs often suffer from {\it catastrophic forgetting}, and thus lose previously acquired knowledge when new information is learned, particularly in dynamic environments requiring continual learning. To tackle this problem, we propose CF-KAN, a new CF method utilizing Kolmogorov-Arnold networks (KANs). By learning nonlinear functions on the \textit{edge level}, KANs are more robust to the catastrophic forgetting problem than MLPs. Built upon a KAN-based autoencoder, CF-KAN is designed in the sense of effectively capturing the intricacies of sparse user--item interactions and retaining information from previous data instances. Despite its simplicity, our extensive experiments demonstrate 1) CF-KAN's superiority over state-of-the-art methods in recommendation accuracy, 2)  CF-KAN's resilience to catastrophic forgetting, underscoring its effectiveness in both static and dynamic recommendation scenarios, and 3) CF-KAN's edge-level interpretation facilitating the explainability of recommendations.
\end{abstract}

%
\section{1. Introduction}

{\bf Background.} Collaborative filtering (CF) is essential in recommender systems, leveraging user--item interactions to provide personalized recommendations. Over time, standard CF techniques have evolved into sophisticated architectures based on multi-layer perceptrons (MLPs), whose main principle involves applying a \textit{fixed} nonlinear activation function to every node in the same layer after a linear transformation. For example, MLP-based autoencoders were leveraged by reconstructing interactions between each user and all items \cite{liang2018variational, wu2016collaborative}. MLPs were utilized for learning the denoising process in diffusion models for CF \cite{wang2023diffusion, hou2024collaborative}. However, MLPs are known to be prone to \textit{catastrophic forgetting}, where the model loses previously acquired knowledge when new information is learned \cite{ramasesh2020anatomy,kemker2018measuring,liu2024kan}, which may lead to suboptimal recommendation accuracy.

Meanwhile, Kolmogorov-Arnold networks (KANs) \cite{liu2024kan} have recently emerged as a promising alternative neural network architecture to MLPs. Inspired by the Kolmogorov-Arnold representation theorem \cite{kolmogorov1961representation}, KANs were designed to overcome fundamental limitations of MLPs \cite{liu2024kan, abueidda2024deepokan,shukla2024comprehensive}. Specifically, unlike MLPs, which have fixed activation functions on \textit{nodes}, KANs contain learnable activation functions on \textit{edges (weights)}. This unique architecture enables KANs to learn nonlinear functions more effectively and to be robust against catastrophic forgetting, making them particularly suited for environments that require continual learning \cite{liu2024kan, herbozo2024kan}. 

{\noindent \bf Motivation.} 
While KANs have generally proven to be highly effective, their performance does not always surpass that of MLPs across all domains. For instance, KANs have shown superior results over MLPs in regression tasks for physics equations \cite{liu2024kan, abueidda2024deepokan}, as well as in time series data \cite{genet2024tkan, vaca2024kolmogorov}. However, in the image domain, KANs may underperform compared to MLPs or convolutional neural networks (CNNs) unless they are carefully designed and optimized \cite{azam2024suitability}. This is because a na\"ive application of KANs falls short of effectively modeling the spatial dependence of local pixels in the image domain \cite{bodner2024convolutional,li2024u}. Likewise, although KANs are powerful, it is crucial to carefully assess their suitability for each domain and appropriately design a particular model to ensure optimal performance. However, the potential of KANs over MLPs in the recommendation domain remains unexplored yet, which motivates us to initiate our study.


\begin{figure}[t]
    \centering
    \includegraphics[width=0.8\columnwidth]{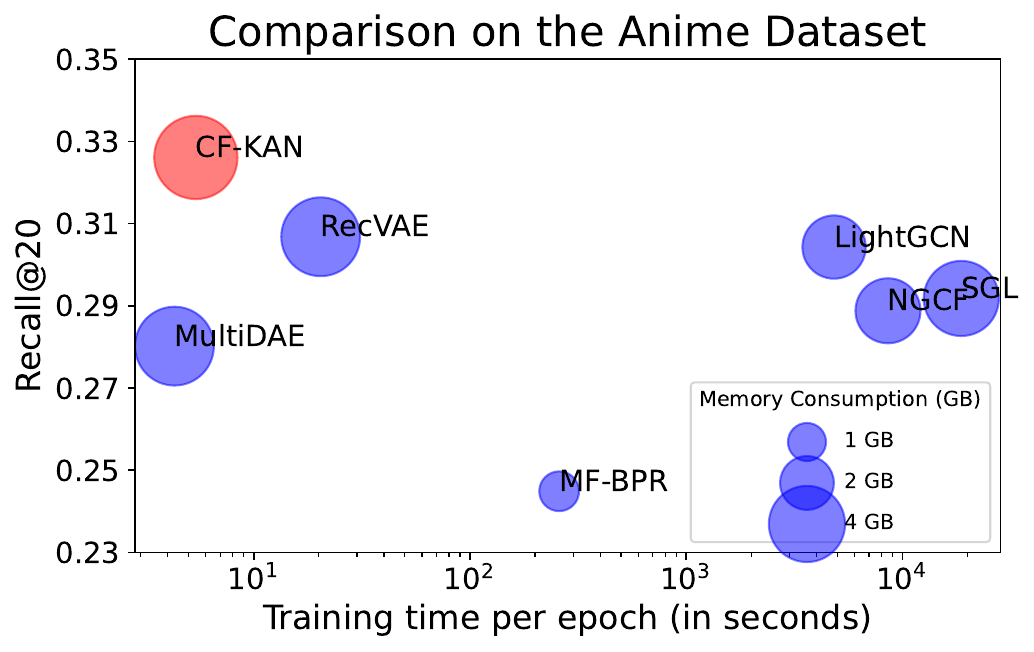}
    \caption{Comparison of CF-KAN and benchmark methods in terms of accuracy, training time, and memory consumption on the Anime dataset.}
    \label{intro_overall}
\end{figure}
{\noindent \bf Main Contributions.} In this study, we introduce CF-KAN,\footnote{For reproducibility, the source code of CF-KAN is available at \underline{\smash{\url{https://github.com/jindeok/CF-KAN}}}.} a new CF method that makes full use of distinguishable characteristics of KANs for CF. The primary objective of our study is to \textbf{uncover and analyze the potential of KANs for recommender systems}. This involves not only assessing overall performance but also evaluating CF-KAN's effectiveness in various perspectives, including 1) {\it dynamic} settings where models are learned incrementally over time, which is feasible in realistic recommendation scenarios, and 2) model {\it interpretability}. Built on a KAN-based autoencoder architecture, CF-KAN is designed to capture complex collaborative signals and retain information effectively from previous user--item interaction instances, thus leading to superior recommendation performance in both static and dynamic settings. Specifically, the \textit{edge-level} learning in KANs results in localized parameter updates, making a KAN-based architecture suitable for modeling the sparse user--item interactions inherent in recommendation environments. Despite its simplicity, extensive experiments demonstrate that CF-KAN consistently outperforms state-of-the-art methods in terms of recommendation accuracy. It is also empirically verified that CF-KAN is resilient to catastrophic forgetting and interpretable. Additionally, thanks to the principle of a simple design based on autoencoders, CF-KAN achieves much faster training time while maintaining superior accuracy, compared to two-tower models \cite{su2023beyond} such as MF-BPR \cite{he2016vbpr} and LightGCN \cite{he2020lightgcn}, which employ separate user query and item encoders \cite{su2023beyond} and require excessive pairwise optimization for all existing user--item interactions. Figure \ref{intro_overall} visualizes these advantages of CF-KAN over state-of-the-art methods. 

We summarize the contributions of our paper in threefold:
\begin{enumerate}
    \item \textbf{Methodology}: We propose CF-KAN, a pioneering approach to harnessing unique properties of KANs for developing a CF method. In contrast to traditional MLP-based CF methods, CF-KAN can directly learn and adapt nonlinear functions at the edge level, addressing the issue of catastrophic forgetting.
    
    \item \textbf{Comprehensive analysis}: We systematically carry out comprehensive experiments to validate the superiority of CF-KAN in various perspectives, including recommendation accuracy, robustness in continual learning scenarios, and scalability. Our results demonstrate that CF-KAN 1) consistently outperforms existing state-of-the-art methods by up to 8.2\% in terms of the Recall@20, 2) showcases its outstanding performance over its counterpart ({\it i.e.}, the MLP variant) in dynamic recommendation environments, and 3) exhibits fast training speeds.

    \item \textbf{Enhanced Interpretability}: Extensive case studies via visualizations demonstrate that CF-KAN is fairly interpretable through its edge-specific learning and pruning by highlighting the importance of individual user--item interactions. Such interpretations are vital for model transparency and user confidence. 

\end{enumerate}


\section{2. Methodology}
In this section, we first describe KAN as a preliminary. Next, we elaborate on CF-KAN, our proposed method. Additionally, we scrutinize CF-KAN in terms of internal model behaviors in both continual learning scenarios and its interpretability.

\subsection{2.1. Kolmogorov-Arnorld Network}

Recently, KAN \cite{liu2024kan} has been proven to serve as a promising alternative to MLP. While MLP is based on the universal approximation theorem \cite{cybenko1989approximation}, KAN is grounded in the Kolmogorov-Arnold (KA) representation theorem \cite{kolmogorov1961representation}. 

\begin{theorem}[KA Representation Theorem]
\label{KAtheorem}
Let $f$ be a multivariate continuous function on a bounded domain. Then, $f$ can be represented as a finite composition of two argument addition of continuous functions of a single variable. Specifically, for a smooth function $f : [0, 1]^n \to \mathbb{R}$, it holds that
\begin{equation}
f({\bf x}) = f(x_1, \ldots, x_n) = \sum_{q=1}^{2n+1} \Phi_q \left( \sum_{p=1}^n \phi_{q,p}(x_p) \right),
\end{equation}
where $\phi_{q,p} : [0, 1] \to \mathbb{R}$ and $\Phi_q : \mathbb{R} \to \mathbb{R}$ are continuous functions. 
\end{theorem}
On the other hand, a KAN layer $\Phi$ is given by
\begin{equation}
\label{eq_phi}
\Phi = \{\phi_{q,p}\},
\end{equation}
where $\Phi$ is the function matrix and $\phi_{q,p}(x_p)$'s are \textit{learnable} activation functions, where $p = 1,2, \cdots, n_{\text{in}}$ and $q = 1,2, \cdots, n_{\text{out}}$; and $n_{\text{in}}$ and $n_{\text{out}}$ are the input and output dimensions of each KAN layer. According to Theorem \ref{KAtheorem}, the inner functions constitute a KAN layer with $n_{in} = n$ and $n_{out} = 2n+1$, while the outer functions form a KAN layer with $n_{in} = 2n+1$ and $n_{out} = n$. Therefore, the KA representations are essentially compositions of these two KAN layers \cite{liu2024kan}. Finally, generalizing the architecture to arbitrary depths and widths leads to an $L$-layer KAN, which is formulated as a composition of continuous functions as follows:
\begin{equation}
\text{KAN}({\bf x}) = f({\bf x}) = \Phi_{L-1} \circ \Phi_{L-2}\circ  \cdots \circ \Phi_{1} \circ \Phi_{0} ({\bf x}).
\end{equation}


\begin{figure}[t]
    \centering
    \includegraphics[width=1\columnwidth]{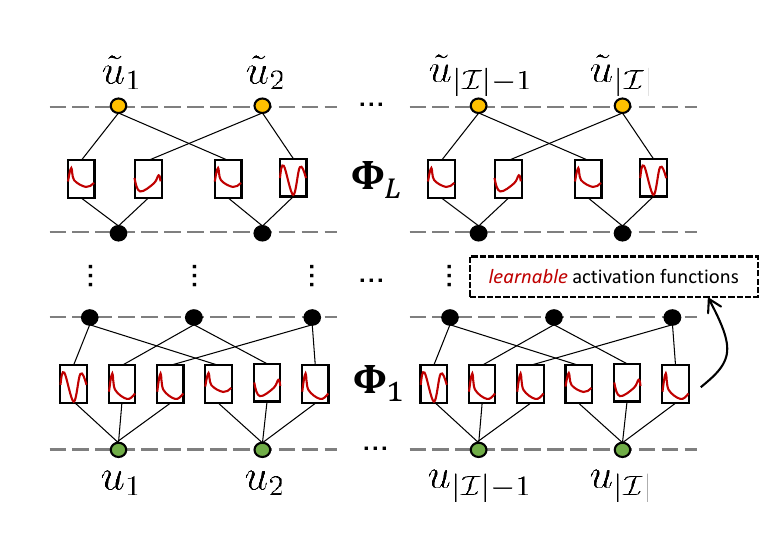}
    \vspace{-5mm}
    \caption{The schematic overview of CF-KAN.}
    \label{overview}
\end{figure}


\subsection{2.2. CF-KAN}

In this subsection, we first describe the model architecture and optimization of CF-KAN. Next, we analyze the unique characteristics of CF-KAN, including its effectiveness in continual learning scenarios and its interpretability. The schematic overview of CF-KAN is illustrated in Figure \ref{overview}.
\subsubsection{Notation.}
We begin by defining the notations. Let $u \in \mathcal{U}$ and $v \in \mathcal{I}$ denote a user and an item, respectively, where $\mathcal{U}$ and $\mathcal{I}$ denote the sets of all users and all items, respectively. The historical interactions of a user $u \in \mathcal{U}$ with items are represented by a binary vector ${\bf u} \in \left\{ 0,1 \right\}^{\left| \mathcal{I} \right|}$, where the $v$-th entry is 1 if there is implicit feedback (such as a click or a view) between user $u$ and item $v \in \mathcal{I}$, and 0 otherwise.

\subsubsection{Model architecture.}
KAN differs from MLP in that, instead of applying the same activation function to all nodes within a layer, KAN learns different nonlinearities at the edge level. This allows KAN to learn nonlinearities more adaptively for each dimension of the input vector (\textit{i.e.}, each item). Thus, the parameters in KAN are likely to be updated {\it locally} during the learning process \cite{liu2024kan}. Meanwhile, in recommender systems, user--item interactions are often extremely sparse, which means that important interaction data are scattered and concentrated in a few key areas rather than being uniformly distributed. In this context, we argue that KAN is more suitable for modeling such intrinsic user--item interactions compared to MLP, which updates its weights more globally than KAN. 

According to the design principle of simplicity and interpretability, CF-KAN is designed using an autoencoder architecture, while the input and output dimensions are the same as the number of items, \textit{i.e.}, $\mathbb{R}^{\left| \mathcal{I} \right|}$. This approach contrasts with two-tower models, such as matrix factorization-based and GCN-based methods, which can accompany high computational costs with an increased number of interactions and often limit interpretations due to predictions via the inner-product of user and item embeddings. By employing an autoencoder, CF-KAN is capable of overcoming these issues, ensuring efficient training while preserving the interpretability of KAN. The autoencoder-based CF-KAN consists of an encoder that maps the input vector ${\bf u}$ to a latent space and a decoder that reconstructs the input from the latent representation: 
\begin{equation}
\label{eq:encdec}
\begin{aligned}
    f_{\text{enc}}({\bf u}) = {\bf z}; f_{\text{dec}}({\bf z}) = \tilde{\bf u},
\end{aligned}
\end{equation}
where $f_{\text{enc}}(\cdot)$ and $f_{\text{dec}}(\cdot)$ are the KAN encoder and KAN decoder, respectively, per layer in CF-KAN; ${\bf z} \in\mathbb{R}^{h}$ represents the $h$-dimensional latent representation obtained from $f_{\text{enc}}(\cdot)$; and $\tilde{\bf u}\in\mathbb{R}^{|\mathcal{I}|}$ is the reconstructed output (\textit{i.e.}, the predicted preference of user $u$) from $f_{\text{dec}}(\cdot)$. More precisely, in CF-KAN, the encoder first maps the input vector ${\bf u}$ to latent representations ${\bf z}$ by directly applying the composition of activation functions $f_{\text{enc}}(\cdot)$, and the decoder $f_{\text{dec}}(\cdot)$ reconstructs the input from ${\bf z}$, which can be formally expressed as
\begin{equation}
\begin{aligned}
    {\bf z} &= f_{\text{enc}}({\bf u}) = \Phi^{(e)}_{E-1} \circ \Phi^{(e)}_{E-2} \circ \cdots \circ \Phi^{(e)}_{1} \circ \Phi^{(e)}_{0} ({\bf u}); \\
    {\tilde{\bf u}} &= f_{\text{dec}}({\bf z}) = \Phi^{(d)}_{D-1} \circ \Phi^{(d)}_{D-2} \circ \cdots \circ \Phi^{(d)}_{1} \circ \Phi^{(d)}_{0} ({\bf z}), 
\end{aligned}
\end{equation}
where $E$ and $D$ are the number of KAN layers in $f_{\text{enc}}(\cdot)$ and $f_{\text{dec}}(\cdot)$, respectively. Here, each $\Phi$ in $f_{\text{enc}}(\cdot)$ and $f_{\text{enc}}(\cdot)$ consists of \textit{learnable} activation functions $\phi_{q,p}: [0,1] \rightarrow \mathbb{R}$, as clearly described in Eq. \eqref{eq_phi}.\footnote{To simplify notations, $\Phi_l^{(\cdot)}$ will be written as $\Phi$ if dropping the subscript and superscript does not cause any confusion.} A practical choice of $\phi_{q,p}$ involves including a basis function such that the activation function $\phi_{q,p}$ is the sum of basis functions \cite{liu2024kan}. Given $u_p$ that represents the $p$-th component of $\bf u$, we formulate $\phi_{q,p}$ as follows:
\begin{equation}
\label{phi_eq}
\phi_{q,p}(u_p) = w_p(\sigma(u_p) + \text{spline}(u_p)),
\end{equation}
where $w_p$ is a learnable parameter; $\sigma(\cdot)$ is an activation function such as PReLU \cite{he2015delving}, ELU \cite{clevert2015fast}, and SiLU \cite{elfwing2018sigmoid}; and $\text{spline}(u_p)$ is expressed as a linear combination of B-splines \cite{piegl1995b}:
\begin{equation}
\label{spline_eq}
 \quad \text{spline}(u_p) = \sum^{G+k-1}_{i=0} c_i B_i(u_p),
\end{equation}
with $c_i$ being learnable parameters, which indicate the position of the control point in the spline; $B_i$ is the $i$-th B-spline base; $k$ is the order of B-spline, which is usually set to $3$ by default \cite{liu2024kan}; and $G$ is the number of grids in the KAN layer. To initialize learnable parameters $w_p$ and $c_i$, we use He initialization \cite{he2015delving}, unlike the original KAN implementation \cite{liu2024kan}.

\subsubsection{Optimization.}

CF-KAN is trained in the sense of minimizing the reconstruction error between the original input ${\bf u}$ and its reconstructed counterpart $\tilde{\bf u}$. The reconstruction error is defined using a loss function $\ell(\cdot)$, such as the mean squared error (MSE) or cross-entropy loss, as in \cite{wu2016collaborative}. The objective function can be formulated as:
\begin{equation}
\mathcal{L} = \frac{1}{|\mathcal{U}|} \sum_{u \in \mathcal{U}} \ell({\bf u}, \tilde{\bf u}) + \lambda \Omega({\bf \Phi}_\text{set}),
\end{equation}
where ${\bf \Phi}_\text{set}$ is the set of all $\Phi$'s over the KAN layers in CF-KAN; $\Omega({\bf \Phi}_\text{set})$ is the regularization term to prevent overfitting and to promote sparsification; and $\lambda$ is the regularization coefficient. Specifically, the regularization term of $\Phi$ is defined as the sum of the $L_1$-norm of the activation function, $|\Phi|_1$, and the additional entropy regularization $S(\Phi)$ \cite{liu2024kan}:
\begin{align}
\label{reg_term}
\Omega({\bf \Phi}_\text{set}) = \sum_{\Phi \in {\bf \Phi}_\text{set}}(|\Phi|_1 + S(|\Phi|)),
\end{align}
where 
\begin{align}
    |\Phi|_1 &= \sum_{p=1}^{n_{\text{in}}}\sum_{q=1}^{n_{\text{out}}}|\phi_{q,p}|_1, \\
    S(\Phi) &= -\sum_{p=1}^{n_{\text{in}}}\sum_{q=1}^{n_{\text{out}}}  {\frac{|\phi_{q,p}|_1}{|\Phi|_1} \log\left(\frac{|\phi_{q,p}|_1}{|\Phi|_1}\right)}
\end{align}
for each KAN layer with $n_\text{in}$ input and $n_\text{out}$ output dimensions.



\begin{figure}[t]
    \centering
    \includegraphics[width=1\columnwidth]{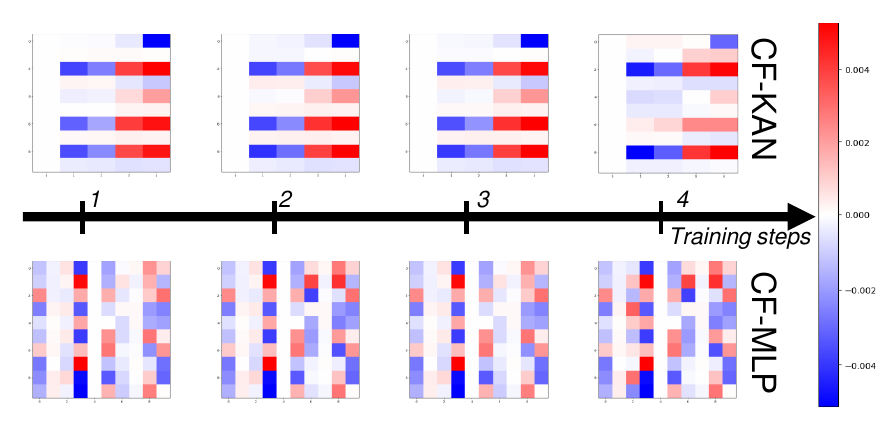}
    \caption{Heatmap visualization of model parameter variations for both CF-KAN and CF-MLP over the training steps on the MovieLens-1M dataset, when $h$ is set to 10 and 10 items are sampled for visualization. Each entry \( (q,p) \) in the heatmap represents the variation of parameters $c_0$'s in $\phi_{q,p}$ of CF-KAN and the variation of parameters in the weight matrix of CF-MLP.
}
    \label{locality}
    \vspace{-0.5mm}
\end{figure}
\vspace{-1mm}
\subsubsection{Application to continual learning.}

The concept of continual learning, where models incrementally learn from a stream of data while retaining previously acquired knowledge, is becoming increasingly important in recommender systems \cite{do2023continual,lee2024continual}. Standard MLP-based models often suffer from catastrophic forgetting, where new data overwrite previously learned knowledge, leading to performance degradation \cite{ ramasesh2020anatomy,kemker2018measuring}. On the other hand, KANs are known to update model parameters more locally and sparsely than MLPs \cite{liu2024kan}, allowing KANs to integrate new data relatively effectively without disrupting previously learned patterns. Our CF-KAN is designed in the sense of leveraging such inherent characteristics of KANs by addressing the challenges of {\it continual learning} in recommender systems. Figure \ref{locality} shows heatmaps visualizing how model parameters in the first layer of both CF-KAN and CF-MLP  change over the training steps on the MovieLens-1M dataset.\footnote{Here, CF-MLP is our MLP variant, where KAN layers in CF-KAN are straightforwardly replaced by MLP layers.} Specifically, for brevity, we visualize how the learnable parameters \( c_0 \) in the matrix \(\Phi_0^{(e)}\) of CF-KAN as well as the weight matrix in the first encoder layer of CF-MLP vary over time. It demonstrates that, for each training step, the model parameters change more globally in CF-MLP than in CF-KAN. 
This implies that CF-KAN has strong potential to maintain relatively high recommendation accuracy in dynamic conditions (\textit{i.e.}, high stability) while adapting well to new information (\textit{i.e.}, high plasticity). 


    \vspace{-0.5mm}

\subsubsection{Interpretability.}
KANs provide nonlinearity specific to each \textit{edge}, making them more interpretable than traditional MLPs \cite{liu2024kan}. Additionally, KANs are likely to be locally updated and sparsely trained with the regularization term in Eq. \eqref{reg_term}, which facilitates pruning while focusing on important edges and nodes.
This motivates us to deal with interpretations based on {\it pruning}, which involves both edge-level and node-level pruning processes. Specifically, for the $l$-th KAN layer of CF-KAN, the node-level pruning drops node $p$ if both incoming edge score $\text{max}_r|\phi_{r,p}|_1$ for layer $l-1$ and outgoing edge score $\text{max}_q|\phi_{q,p}|_1$ for layer $l+1$ are less than threshold $\tau_1$. Similarly, the edge-level pruning eliminates edge $\phi_{r,p}$ if $|\phi_{r,p}|_1$ is less than threshold $\tau_2$.\footnote{Unlike regression tasks for physics equations in \cite{liu2024kan}, the functions to be learned in CF-KAN designed for recommender systems do not necessarily have to be expressed as pre-defined ones such as $e^x$ and $\sin x$. Thus, we omit the symbolification process, simplifying the interpretation process.}

\begin{figure}[t]
    \centering
    \includegraphics[width=1\columnwidth]{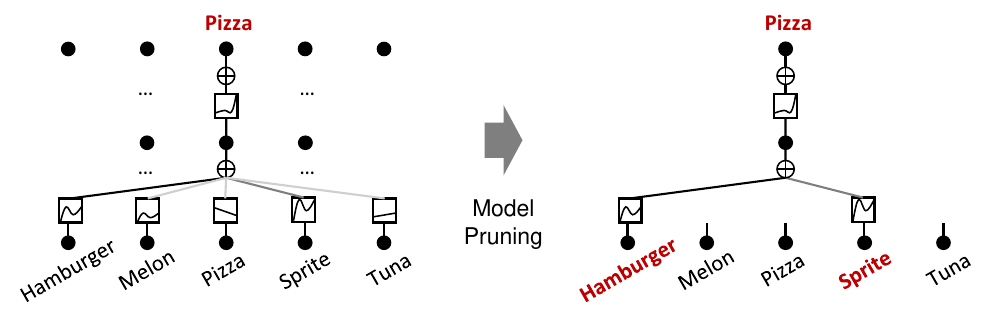}
    \caption{A toy example of interpretations where CF-KAN explains why pizza is recommended to a given user based on his/her past consumption ({\it i.e.}, hamburger and Sprite).}
    \label{interpret_food}
\end{figure}

 The \textit{edge-specific learning} in CF-KAN helps highlight the importance of individual user--item interactions, enabling us to understand the model's internal behavior. For recommender systems, such interpretability of CF-KAN is particularly valuable. 
 Figure \ref{interpret_food} illustrates an example of interpretations where the pruned KAN can produce a proper explanation of why pizza is recommended to a given user based on his/her past consumption ({\it i.e.},  hamburger and Sprite, rather than melon and tuna). This interpretation aids in boosting market sales for e-commerce platforms by providing clearer insights into recommendation mechanisms.

\begin{table}[t]
\captionsetup{skip=2pt}
\footnotesize
\centering
\begin{tabular}{ccccc}
\toprule
Dataset & $|\mathcal{U}|$ & $|\mathcal{I}|$ & \# of interactions & Density \\
\midrule
ML-1M & 5,949 & 2,810 & 571,531 & 0.0342 \\
Yelp & 54,574 & 34,395 & 1,402,736 & 0.0007 \\
Anime & 73,515 & 11,200 & 7,813,737 & 0.0095 \\
\bottomrule
\end{tabular}
\caption{The statistics of three benchmark datasets.}
\label{table:datasets}
\end{table}
\begin{table*}[t]
\centering
\small
\setlength{\tabcolsep}{4pt}

\begin{tabular}{c|c||ccccc|c}
\hline
\textbf{Method} & \textbf{Metric} & \textbf{After D1} & \textbf{After D2} & \textbf{After D3} & \textbf{After D4} & \textbf{After D5} & \textbf{Avg. gain (\%)} \\ 
\hline
\multirow{3}{*}{\textbf{CF-KAN}} 
& \textbf{LA} & 0.0182 & 0.0184 & 0.0182 & 0.0190 & 0.0192 & \textbf{+4.5} \% \\ 
& \textbf{RA} & 0.0136 & 0.0141 & 0.0103 & 0.0106 & 0.0094 & \textbf{+31.3} \%\\ 
& \textbf{H-mean} & 0.0155 & 0.0159 & 0.0131 & 0.0136 & 0.0127 &\textbf{ +20.8} \%\\ 
\hline
\multirow{3}{*}{\textbf{CF-MLP}} 
& \textbf{LA} & 0.0175 & 0.0177 & 0.0177 & 0.0180 & 0.0181 & - \\ 
& \textbf{RA} & 0.0131 & 0.0082 & 0.0095 & 0.0064 & 0.0088 & - \\ 
& \textbf{H-mean} & 0.0150 & 0.0112 & 0.0123 & 0.0094 & 0.0118 & - \\ 
\hline
\end{tabular}
\caption{Performance comparison of CF-KAN and CF-MLP in terms of LA, RA, and H-mean scores when R@20 is adopted on the ML-1M dataset.}
\label{tab:cont}
\vspace{-2mm}
\end{table*}
\section{3. Experiments}

In this section, we conduct comprehensive experiments that are designed to answer the following five key research questions (RQs):
\begin{itemize}
    \item \textbf{RQ1:} How robust is KAN against catastrophic forgetting in continual learning scenarios compared to MLP?
    \item \textbf{RQ2:} How much does CF-KAN improve the top-$K$ recommendation accuracy over benchmark CF methods?
    \item {\bf RQ3}: How interpretable is CF-KAN?
    \item {\bf RQ4}: How scalable is CF-KAN in terms of both training time and consumed memory?
    \item {\bf RQ5}: How sensitive is the performance of KAN to its key parameters?
\end{itemize}

\subsection{3.1. Experimental Settings}
\subsubsection{Datasets.} 
We perform our experiments using three widely used real-world datasets, namely MovieLens-1M (ML-1M), Yelp, and one {\it large-scale} dataset, Anime \cite{wang2023diffusion,he2020lightgcn,hou2024collaborative,wang2019neural}. A summary of the statistics for each dataset is summarized in Table \ref{table:datasets}.

\subsubsection{Evaluation protocol.} We adopt two widely used top-$K$ ranking metrics, Recall@$K$ (R@$K$) and NDCG@$K$ (N@$K$), where $K \in \left\{ {10,20} \right\} $. Especially for the evaluation of continual learning, we use three standard metrics (the higher the better), including the learning average (LA), retained average (RA), and H-means \cite{do2023continual,lee2024continual}. Specifically, given $a_{i,j}$ which representing the recommendation performance on block $j$ after training on block $i$, the three metrics are computed as follows:
\begin{itemize}
    \item \textbf{LA}: $\frac{1}{k}\sum^k_{i=1}a_{i,i}$. This metric evaluates how effectively a model adapts to new data blocks, measuring plasticity.
    \item \textbf{RA}: $\frac{1}{k}\sum^k_{i=1}a_{k,i}$. This metric evaluates how well a model retains previously learned knowledge, measuring stability.
    \item \textbf{H-mean}: This metric is the harmonic mean of LA and RA.
\end{itemize}

\subsubsection{Competitors.}
To demonstrate the superiority of CF-KAN, we comprehensively compare its recommendation accuracy against thirteen benchmark CF methods employing four different base model architectures as follows. 
\begin{itemize}
\item \textbf{Matrix factorization-based methods}: MF-BPR \cite{rendle2012bpr}, NeuMF \cite{he2017neural}, and DMF \cite{xue2017deep};
\item \textbf{Autoencoder-based methods}: CDAE \cite{wu2016collaborative}, Multi-DAE \cite{liang2018variational}, and RecVAE \cite{shenbin2020recvae};
\item \textbf{GCN-based methods}: SpectralCF \cite{zheng2018spectral}, NGCF \cite{wang2019neural}, LightGCN \cite{he2020lightgcn}, SGL \cite{wu2021self}, and NCL \cite{lin2022improving};
\item \textbf{Generative model-based methods}: CFGAN \cite{chae2018cfgan}, RecVAE \cite{shenbin2020recvae}, and DiffRec \cite{wang2023diffusion}.
\end{itemize}

\subsubsection{Implementation details.} We use the best hyperparameters for both competitors and CF-KAN, determined through hyperparameter tuning on the validation set. We search for the hyperparameters in the following ranges: $\left\{ {1,2,3,4,5} \right\} $ for the number of grids, $G$; $\left\{ {128,256,512,1024} \right\} $ for the latent dimension $d$; $\left\{ {1,2,3,4} \right\} $ for the number of layers in the encoder and decoder, $E=D=L$. We use the Adam optimizer \cite{kingma2015adam}, where the batch
size is set to 256 for all experiments. For the choice of $l(\cdot)$, we use the MSE loss on ML-1M and Anime, and the cross-entropy loss on Yelp. We use the spline order $k$ of 3 by default. For baseline implementation, we use Recbole \cite{zhao2021recbole}, an open-sourced recommendation framework. All experiments are conducted on a machine with Intel (R) 12-Core (TM) i7-9700K CPUs @ 3.60 GHz and an NVIDIA GeForce RTX A6000 GPU.

\begin{table*}[!t]\centering
\setlength\tabcolsep{5.0pt}

  \captionsetup{skip=2.0pt}
  \small
  \begin{tabular}{c|cccc|cccc|cccc}
    \toprule[1pt]
    \multicolumn{1}{c|}{}&\multicolumn{4}{|c|}{ML-1M}&\multicolumn{4}{c|}{Yelp}&\multicolumn{4}{c}{Anime}\\
    \cmidrule{1-13}
           Method & R@10& R@20& N@10& N@20& R@10& R@20& N@10& N@20& R@10& R@20& N@10& N@20\\
    \midrule[1pt]
    MF-BPR & 0.0876& 0.1503& 0.0749&0.0966& 0.0341& 0.0560& 0.0210& 0.0341 & 0.1521& 0.2449 & 0.2925 & 0.3153\\
    NeuMF & 0.0845& 0.1465& 0.0759& 0.0965&  0.0378& 0.0637& 0.0230& 0.0308&0.1531& 0.2442&0.3277& 0.3259\\
    DMF & 0.0799& 0.1368& 0.0731& 0.0921&  0.0342& 0.0588& 0.0208& 0.0282&0.1386& 0.2161&0.3277& 0.3122\\
    \cmidrule{1-13}
    
    
    CDAE & 0.0991& 0.1705& 0.0829&0.1078& 0.0444& 0.0703& 0.0280& 0.0360 & 0.2031& 0.2845 & 0.4652 & 0.4301\\
    MultiDAE& 0.0975& 0.1707& 0.0820& 0.1046& 0.0531& 0.0876& 0.0316& 0.0421 & 0.2022& 0.2802 & 0.4577 & 0.4125\\
    RecVAE& 0.0835& 0.1422& 0.0769& 0.0963& 0.0493& 0.0824& 0.0303& 0.0403 & \underline{0.2137} & \underline{0.3068} & 0.4105 & 0.4068\\
    \cmidrule{1-13}
    SpectralCF & 0.0751& 0.1291& 0.0740&0.0909& 0.0368& 0.0572& 0.0201& 0.0298 & 0.1633& 0.2564 & 0.3102 & 0.3236\\
    NGCF & 0.0864& 0.1484& 0.0805&0.1008& 0.0428& 0.0726& 0.0255& 0.0345 & 0.1924& 0.2888 & 0.3515 & 0.3485\\
    LightGCN& 0.0824& 0.1419& 0.0793& 0.0982& 0.0505& 0.0858& 0.0312& 0.0417 & 0.2071 & 0.3043 & 0.3937 & 0.3824\\
    SGL& 0.0885& 0.1575& 0.0802& 0.1029& \underline{0.0564}& \underline{0.0944}& \underline{0.0346}& \underline{0.0462} & 0.1994 & 0.2918 & 0.3748 & 0.3652\\
    NCL& 0.0878& 0.1471& 0.0819& 0.1011& 0.0535& 0.0906& 0.0326& 0.0438 & 0.2063 & 0.3047 & 0.3915 & 0.3819\\
    \cmidrule{1-13}
    CFGAN& 0.0684& 0.1181& 0.0663& 0.0828& 0.0206& 0.0347 & 0.0129 &0.0172 & 0.1946 & 0.2889& 0.4601& 0.4289\\       
    DiffRec& \underline{0.1021}& \underline{0.1763}& \underline{0.0877}& \underline{0.1131}& 0.0554& 0.0914& 0.0343& 0.0452 & 0.2104 & 0.3012 & \underline{0.5047} & \underline{0.4649}\\
    \cmidrule{1-13}
    \textbf{CF-KAN}& {\bf 0.1065}& {\bf 0.1831}& {\bf 0.0894}& {\bf 0.1152}& {\bf 0.0594}& {\bf 0.0974}& \bf{0.0363}& \bf{0.0478} &{\bf 0.2287} & {\bf 0.3261} & {\bf 0.5256} & {\bf 0.4875}\\
    \bottomrule[1pt]
  \end{tabular}
  \caption{Performance comparison among CF-KAN and thirteen recommendation competitors for the three benchmark datasets. Here, the best and second-best performers are highlighted by bold and underline, respectively. The improvements of CF-KAN over the best competitors are all statistically significant with $p$-value $\leq 0.01$.}
\label{tab:comparison}
\vspace{-2mm}
\end{table*}

\subsection{3.2. RQ1: Continual Learning Scenarios}

\begin{figure}[t]
 
        \centering
        \begin{subfigure}[c]{0.47\columnwidth}
                \includegraphics[width=1\columnwidth]{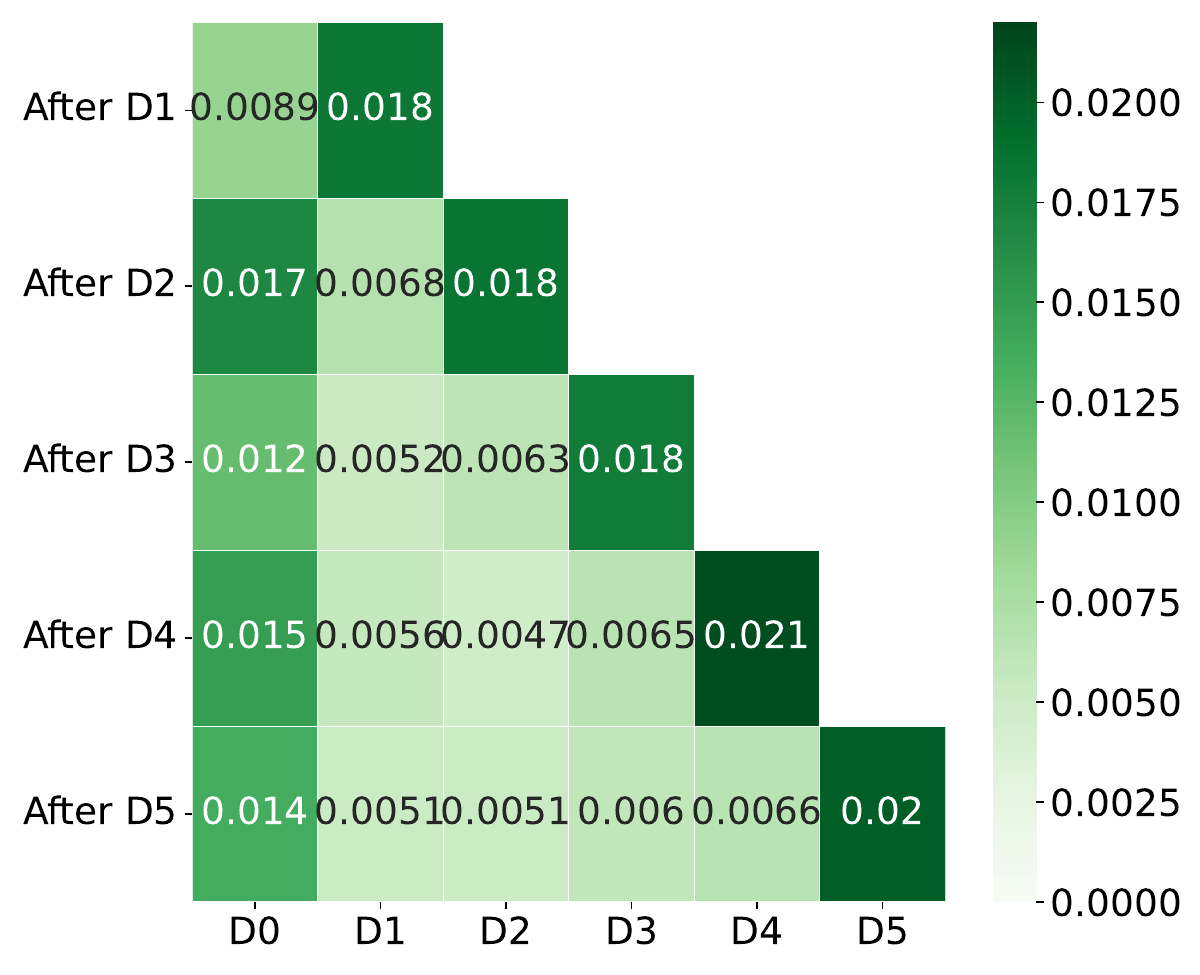}
                \caption{R@20, CF-KAN}
        \end{subfigure}        
        \begin{subfigure}[c]{0.47\columnwidth}
                \includegraphics[width=1\columnwidth]{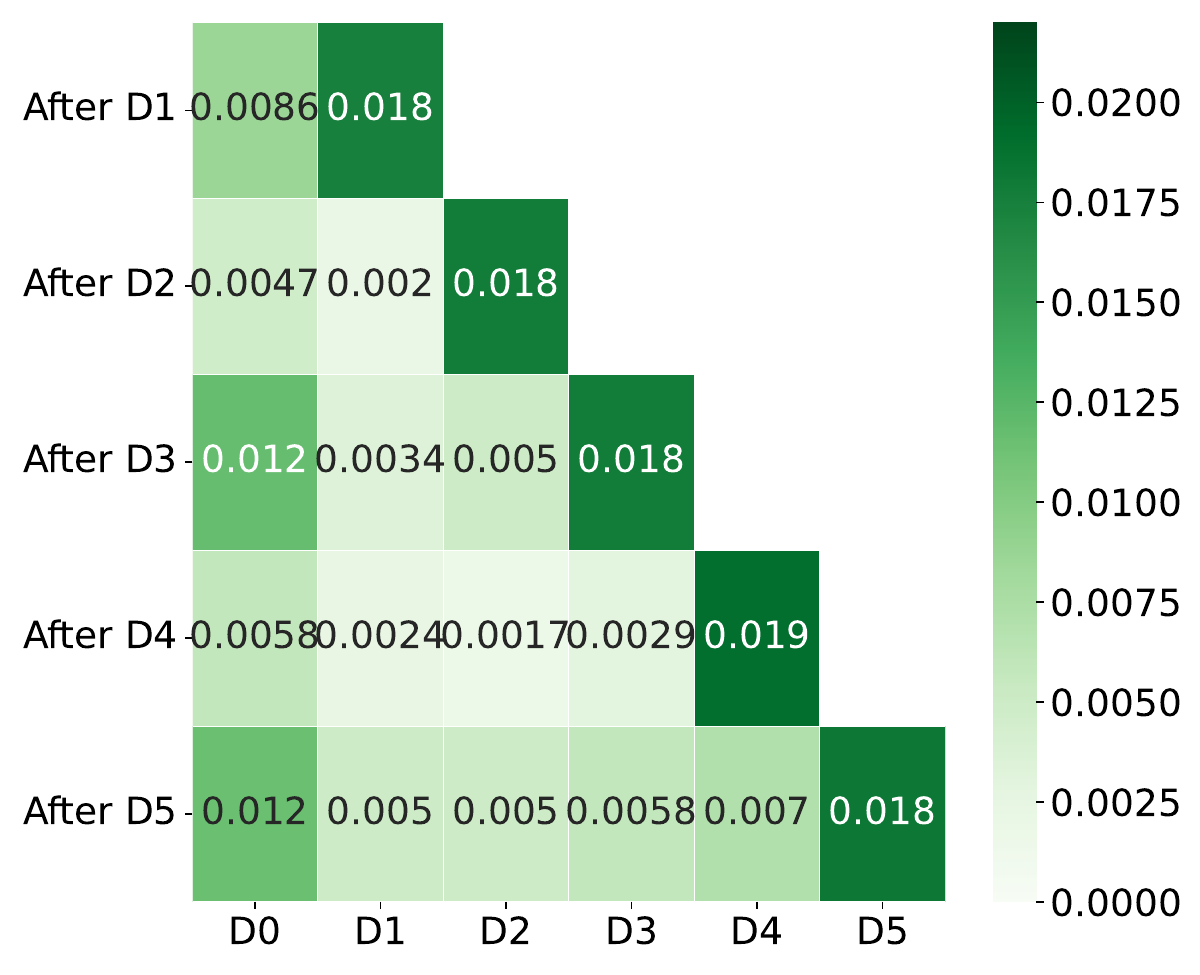}
                \caption{R@20, CF-MLP}
        \end{subfigure}
        \caption{Performance comparison of CF-KAN and CF-MLP in terms of R@20 on the ML-1M dataset in the continual learning scenario.}
\label{fig:cont_heatmap} 
\end{figure}

As reported in \cite{liu2024kan}, KAN has better ability in continual learning scenarios, exhibiting robustness against the problems of catastrophic forgetting. To empirically validate this, we use the ML-1M dataset where {\it timestamps} are available over the entire interactions. For empirical evaluation, we adhere to the standard continual learning protocols established in \cite{lee2024continual,xu2020graphsail,mi2020ader}. Specifically, each dataset is split in such a way that 50\% constitutes the base block $D_0$, while the remaining 50\% is evenly divided into 5 incremental blocks ($D_1, \cdots, D_5$) based on timestamps. Within each block, interactions are further divided into training, validation, and test sets in a ratio of 80\%/10\%/10\%. To observe apparent gains of KAN against MLP in the recommendation domain, we compare CF-KAN with CF-MLP, which is an MLP variant of CF-KAN where KAN is straightforwardly replaced by MLP.\footnote{For a fair comparison, we set the number of parameters of both models identically. Note that CF-MLP can be regarded as Multi-DAE \cite{liang2018variational}.} Our empirical findings are as follows:

\begin{enumerate}[label=(\roman*)]
\item As shown in Figure \ref{fig:cont_heatmap} and Table \ref{tab:cont}, CF-KAN consistently outperforms CF-MLP for all the metrics across all time points. This demonstrates that CF-KAN surpasses CF-MLP in terms of both plasticity and stability, effectively adapting to new data while retaining previously learned information.
\item The results in Table \ref{tab:cont} indicate that gains in RA over CF-MLP are indeed significant. This implies that CF-KAN is markedly superior to CF-MLP in the sense of retaining past knowledge. 
\end{enumerate}

\begin{figure}[t]
        \centering
        \begin{subfigure}[c]{1\columnwidth}
                \includegraphics[width=0.95\columnwidth]{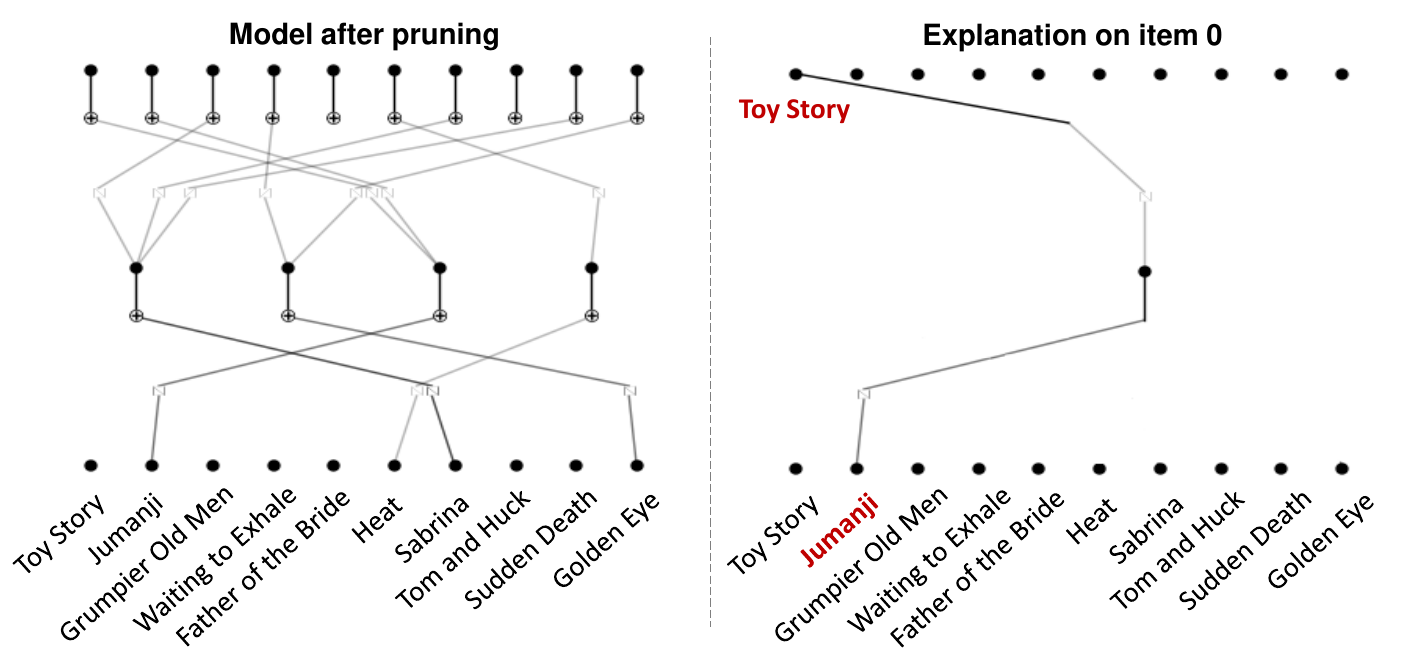}
                \caption{ML-1M}
                \label{fig:ml1m_exp}
        \end{subfigure}        
        \begin{subfigure}[c]{1\columnwidth}
                \includegraphics[width=0.95\columnwidth]{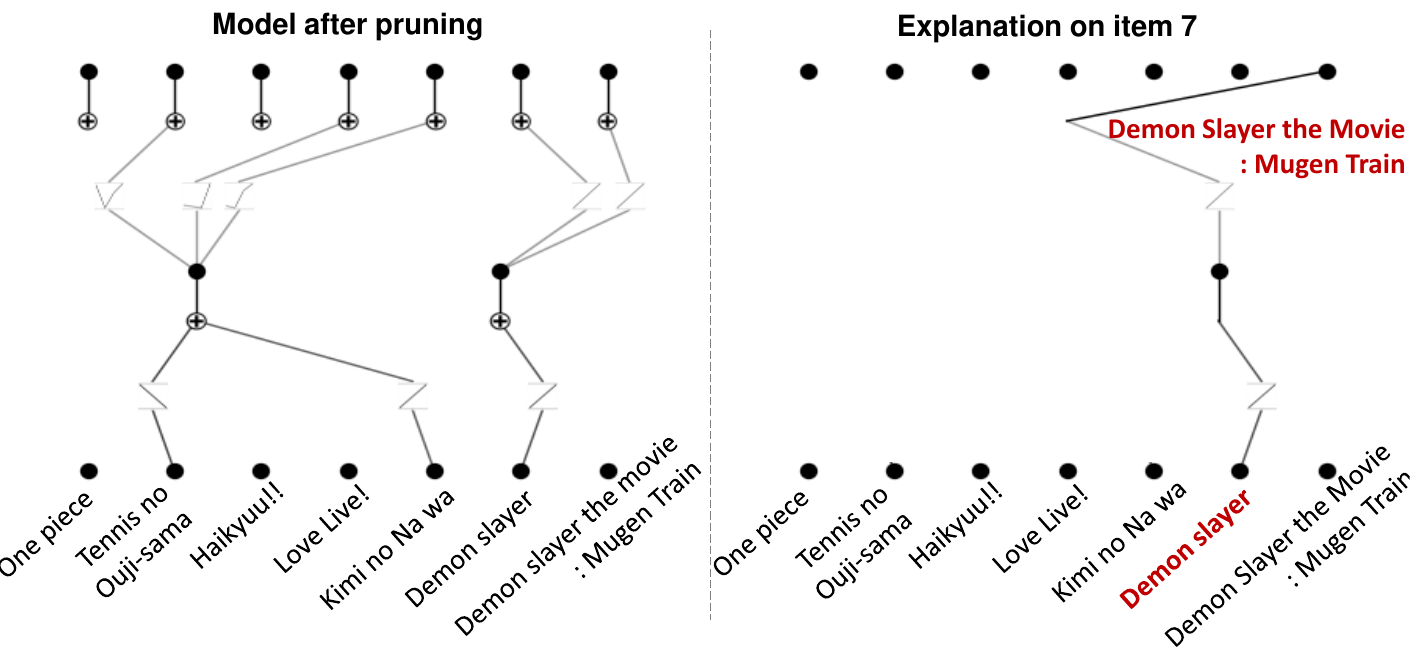}
                \caption{Anime}
                \label{fig:anime_exp}
        \end{subfigure}
        \caption{Visualization of CF-KAN's interpretation on the (a) ML-1M and (b) Anime datasets. The left panel shows the model after training and pruning, and the right panel shows the explanation of a particular item recommendation.}
        \label{fig:interpret} 
\end{figure}

\subsection{3.3. RQ2: Recommendation Accuracy}

To assess the recommendation accuracy of CF-KAN, we conducted extensive experiments on the three benchmark real-world datasets: ML-1M, Yelp, and Anime. From Table \ref{tab:comparison}, our findings are as follows:

\begin{enumerate}[label=(\roman*)]
\item Even with a simple KAN-based autoencoder architecture, CF-KAN achieves state-of-the-art performance on the three datasets with gains up to 8.2\% over the best competitors, due to its superior capability to capture nonlinear relationships within complex user--item interactions ({\it i.e.}, collaborative signals).
\item CF-KAN outperforms CF methods employing rather sophisticated models such as variational autoencoders (RecVAE), GCNs (NGCF and LightGCN), and diffusion-based models (DiffRec). 
\item Particularly noteworthy is the superior performance of CF-KAN over MLP variants such as CDAE and Multi-DAE. This highlights the potential of KANs, in that KANs are more adept at learning functions in CF than MLPs.
\end{enumerate}

\subsection{3.3. RQ3: Interpretation} \label{sec:interpret}

We conduct case studies on interpretations using real-world datasets. Figures \ref{fig:ml1m_exp} and \ref{fig:anime_exp} illustrate the model interpretability on the ML-1M and Anime datasets (for movie/anime recommendations), respectively. For visualization, a subset of items from each dataset was sampled and utilized during training. The left panels in Figure \ref{fig:interpret} show the results after the models were trained and pruned with thresholds $\tau_1$ and $\tau_2$ set to $1e^{-1}$ and $9e^{-2}$, respectively. Here, the thickness of each connected edge reflects the relative importance score $|\phi_{q,p}|_1$, and the learned function on each edge is visualized.

By tracing the edges connected to the target output item back to the input layer, we can identify which items significantly influence the recommendation of a particular item. For example, in Figure \ref{fig:ml1m_exp}, tracing the edges connected to `\textit{Toy Story}' reveals that watching `\textit{Jumanji}' strongly influences the model's high recommendation probability for `\textit{Toy Story}'. Given that `\textit{Toy Story}' and `\textit{Jumanji}' belong to similar genres, such as Children and Adventure, the validity of the explanation is confirmed. Similarly, in Figure \ref{fig:anime_exp}, the recommendation of `\textit{Demon Slayer the Movie: Mugen Train}' is strongly influenced by the user's viewing history of the `\textit{Demon Slayer}' anime series. These instances clearly demonstrate the interpretability of CF-KAN, which is particularly valuable in understanding and improving recommendation models. Note that CF-MLP is much less capable of producing apparent connection paths corresponding to interpretations, due to its fixed activation functions and global updates.
\begin{table}[!t]\centering
\setlength\tabcolsep{5.0pt}

  \captionsetup{skip=2.0pt}
  \footnotesize
  \begin{tabular}{c|cc|cc|cc}
  
    \toprule[1pt]
    \multicolumn{1}{c|}{} & \multicolumn{2}{c|}{ML-1M} & \multicolumn{2}{c|}{Yelp} & \multicolumn{2}{c}{Anime} \\
    \cmidrule{1-7}
    Method & s/ep & mem. & s/ep & mem. & s/ep & mem. \\
    \midrule[1pt]
    MF-BPR & 3.6& 1.13& 15.4& 1.59 & 258.5& 1.09\\
    NGCF & 33.2&1.43 & 238.2& 2.76& 8571.4& 2.91\\
    LightGCN & 26.1 & 1.02& 208.3& 2.03& 4828.1& 2.76\\
    SGL & 97.4 & 1.53& 853.1& 3.61 & 18711.6& 3.89\\
    \cmidrule{1-7}
    MultiDAE & 0.1& 1.32 & 7.1 &2.52 & 4.3 & 4.28\\
    RecVAE & 0.4& 1.29& 30.2& 2.54& 20.4&4.28 \\
    \textbf{CF-KAN} & 0.2 & 1.42& 14.2& 2.62 & 5.4 & 4.78\\ 
    \bottomrule[1pt]
  \end{tabular}
  \caption{Performance comparison in terms of the training time (in seconds) per epoch and GPU memory (in GB) consumption on the three benchmark datasets. The batch size and embedding dimension are set to $256$ for all experiments.}
\label{tab:scal_comp}
\end{table}
\vspace{-3mm}
\subsection{3.4. RQ4: Scalability Analysis}

To assess the scalability of CF-KAN, we evaluate the training time per epoch and GPU memory consumption across representative two-tower models (MF-BPR, NGCF, LightGCN, and SGL) and autoencoder-based models (MultiDAE, RecVAE, and CF-KAN) on the same experimental settings. From Table \ref{tab:scal_comp}, our observations are as follows:
\begin{enumerate}[label=(\roman*)]
    \item Autoencoder-based methods exhibit significantly faster training speeds compared to two-tower models. This is because two-tower models need to learn all pairwise relationships in proportion to the number of interactions. 
    \item Autoencoder-based methods were found to consume slightly more memory compared to two-tower methods due to the inference requirements across all item dimensions. Nevertheless, on ML-1M and Yelp, CF-KAN even requires less memory than NGCF and SGL.
    \item In comparison with other autoencoder-based methods (MultiDAE and RecVAE), CF-KAN is quite comparable in terms of computational and memory complexities, while revealing higher recommendation accuracy.
\end{enumerate}

\subsection{3.5. RQ5: Sensitivity Analysis}
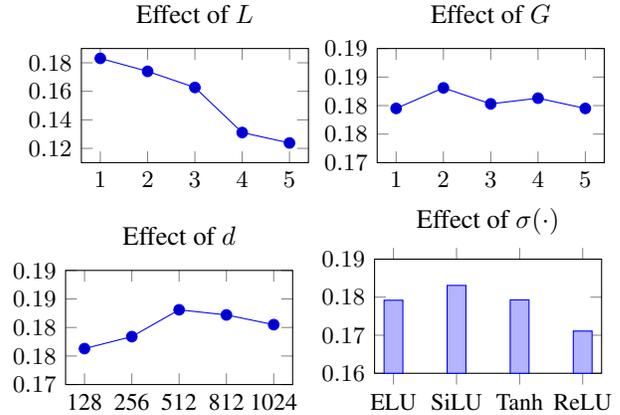
\begin{figure}[t!]
    \centering
    \begin{tikzpicture}
        \begin{axis}[
            width=4.6cm,
            height=3.1cm,
            title={Effect of $L$},
            tick label style={font=\small},
            ymin=0.11, ymax=0.19,
            xtick={1,2,3,4,5}
        ]
        \addplot coordinates {(1,0.1831) (2,0.174) (3,0.1627) (4,0.1311) (5,0.1238)};
        \end{axis}
    \end{tikzpicture}
    \begin{tikzpicture}
        \begin{axis}[
            width=4.6cm,
            height=3.1cm,
            title={Effect of $G$},
            tick label style={font=\small},
            ymin=0.17, ymax=0.19,
            xtick={1,2,3,4,5}
        ]
        \addplot coordinates {(1,0.1795) (2,0.1831) (3,0.1803) (4,0.1813) (5,0.1795)};
        \end{axis}
    \end{tikzpicture}

    \begin{tikzpicture}
        \begin{axis}[
            width=4.6cm,
            height=3.1cm,
            title={Effect of $d$},
            tick label style={font=\small},
            ymin=0.17, ymax=0.19,
            xtick={1,2,3,4,5},
            xticklabels={128, 256, 512, 812, 1024}
        ]
        \addplot coordinates {(1,0.1763) (2,0.1784) (3,0.1831) (4,0.1822) (5,0.1805)};
        \end{axis}
    \end{tikzpicture}
    \begin{tikzpicture}
        \begin{axis}[
            width=4.6cm,
            height=3.1cm,
            ybar,
            symbolic x coords={ ELU, SiLU, Tanh, ReLU},
            xtick=data,
            ymin=0.16, ymax=0.19,
            bar width=0.25cm,
            tick label style={font=\small},
            title={Effect of $\sigma(\cdot)$},
        ]
        \addplot coordinates {(ELU,0.1792) (SiLU,0.1831) (Tanh,0.1793) (ReLU,0.1711)};
        \end{axis}
    \end{tikzpicture}
\caption{Effect of hyperparameters on R@20 for the ML-1M dataset.}
\label{fig:sa} 
\end{figure}

We analyze the impact of key parameters of CF-KAN, including the number of KAN layers ($L$), the number of grids ($G$), the dimension of each hidden layer ($d$), and the choice of activation functions ($\sigma(\cdot)$ in Eq.\eqref{phi_eq}), on the recommendation accuracy using the ML-1M dataset.\footnote{We use the following pivot values for the key parameters: $G=2$, $d=512$, $N=1$, and $\sigma(\cdot)$ is set to $\text{SiLU}$.} From Figure \ref{fig:sa}, our findings are as follows:

\paragraph{(\bf Effect of $L$)} Surprisingly, $L=1$ yields the highest accuracy, and the performance deteriorates as more layers are stacked. This implies that CF-KAN is indeed shallow and the function composition to be learned in CF is simple, which justifies the adoption of a rather simple autoencoder in designing KAN-based CF methods.

\paragraph{(\bf Effect of $G$)} Increasing the number of grids does not necessarily guarantee higher performance. It turns out that using only 2 grids can achieve the optimal performance. This implies that CF-KAN does not require a large number of parameters to effectively learn the model.

\paragraph{(\bf Effect of $d$)} The accuracy tends to be improved as $d$ increases; however, the performance starts to slightly decline beyond $d=812$ due to the overfitting. This underscores the importance of selecting an appropriate value of $d$.

\paragraph{(\bf Effect of $\sigma(\cdot)$)} SiLU achieves the highest performance, while ReLU performs the lowest. This finding coincides with a physics regression task \cite{liu2024kan}.

\section{4. Conclusions and Future Work}

In this study, we explored an open yet important problem of how to overcome the fundamental limit of MLP-based CF techniques experiencing catastrophic forgetting. To this end, we deviced CF-KAN, a new CF method utilizing KANs, which learns nonlinear functions on the edge level and thus is more robust to catastrophic forgetting. Through extensive experiments on various real-world benchmark datasets, we demonstrated that CF-KAN is 1) superior to its counterpart ({\it i.e.}, CF-MLP) in both static and dynamic recommendation environments, 2) highly interpretable via visualizations, and 3) scalable in terms of both training time and memory. Our future research involves exploring the potential of KANs in various recommendation domains. 

\bibliography{citation_list}



\end{document}